\documentclass[journal]{IEEEtran}

\usepackage{fancyhdr}
\fancypagestyle{firstpage}{
  \fancyhf{}
  \fancyhead[C]{This work has been accepted for publication in IEEE Communications Magazine in January 2017. Please, cite the IEEE Communications Magazine.}
}
\usepackage[T1]{fontenc}
\usepackage{color}
\usepackage{amsmath}
\usepackage{graphicx}
\usepackage[cmintegrals]{newtxmath}
\usepackage{epstopdf}
\usepackage{multicol}
\usepackage{multirow}
\usepackage{xspace}

\newcommand{\LoRaWAN}   {LoRaWAN\xspace}
\newcommand{\LoRa}      {LoRa\xspace}

\begin{document}

\title{Understanding the Limits of \LoRaWAN}

\author{
    Ferran~Adelantado,
    Xavier~Vilajosana,
    Pere~Tuset-Peiro,
    Borja~Martinez,
    Joan~Meli\`{a}-Segu\'{i},
    Thomas~Watteyne,
    \thanks{F. Adelantado, P. Tuset-Peiro, B. Martinez and J. Meli\`{a}-Segu\'{i} are with IN3 at the Universitat Oberta de Catalunya.}
    \thanks{X. Vilajosana is with IN3 at Universitat Oberta de Catalunya and Worldsensing.}%
    \thanks{T. Watteyne is with Inria, EVA-team, Paris, France.}
}


\maketitle
\IEEEpeerreviewmaketitle

\thispagestyle{firstpage}

\section*{Abstract}

Low-Power Wide Area Networking (LPWAN) technology offers long-range communication, which enables new types of services.
Several solutions exist; \LoRaWAN is arguable the most adopted.
It promises ubiquitous connectivity in outdoor IoT applications, while keeping network structures, and management, simple.
This technology has received a lot of attention in recent months from network operators and solution providers.
Yet, the technology has limitations that need to be clearly understood to avoid inflated expectations and disillusionment.
This article provides an impartial and fair overview of what the capabilities and the limitations of \LoRaWAN are.
We discuss those in the context of use cases, and list open research and development questions.

\section{Introduction}
\label{sec:introduction}

Network operators are starting to deploy horizontal M2M solutions to cover a wide set of large scale verticals, using Low Power Wide Area Networking (LPWAN) technologies~\cite{linklabs16comprehensive,goursaud16dedicated}.
Application domains include smart city, metering, on-street lighting control or precision agriculture.
LPWAN technologies combine low data rate and robust modulation to achieve multi-km communication range.
This enables simple star network topologies that simplify network deployment and maintenance~\cite{xiong15low}.
While the benefits of these technologies are known and are often considered as the key enablers for some applications, their limitations are still not well understood~\cite{sanchez16state, margelis15low}.

In this article we aim to provide an impartial overview of the limitations of \LoRaWAN~\cite{sornin16lora}, one of the most successful technologies in the LPWAN space.
\LoRaWAN is a network stack rooted in the \LoRa physical layer.
\LoRaWAN features a raw maximum data rate of 27~kbps (50 kbps when using FSK instead of LoRa), and claims that a single gateway can collect data from thousands of nodes deployed kilometers away.
These capabilities have really resonated with some solution providers and network operators, who have created a large momentum behind \LoRaWAN to the point that it is sometimes touted as the connectivity enabler for any IoT use case~\cite{ducrot16lora}.

The goal of this article is to bring some sanity to these statements, by providing a comprehensive, fair and independent analysis of what the capabilities and limitations of \LoRaWAN are.
We adopt a pragmatic approach, and identify in which use cases the technology works, and in which use cases it doesn't work.
Section~\ref{sec:overview} provides an overview of LPWAN technologies, including cellular.
Section~\ref{sec:description} describes \LoRaWAN technology in details.
Section~\ref{sec:capacity} analyzes the network capacity and scale limitations of the technology.
Section~\ref{sec:usecases} discusses the use cases where \LoRaWAN works/doesn't work.
Section~\ref{sec:research} lists open research and development challenges for the technology.
Section~\ref{sec:conclusion} concludes.

\section{Overview of LPWAN and Cellular technologies for IoT}
\label{sec:overview}

\subsection{Low-Power Wide-Area Alternatives}

Although LoRaWAN is one of the most adopted technologies for IoT, there is a wide range of LPWAN technologies in the market, such as Ingenu, Weightless W, N and P or SigFox~\cite{draft-minaburo-lpwan-gap-analysis}.

Ingenu developed a proprietary LPWAN technology in the 2.4~GHz band, based on Random Phase Multiple Access (RPMA) to provide M2M industry solutions and private networks.
The main asset of Ingenu in comparison with alternative solutions is high data rate up to 624~kbps in the uplink, and 156~kbps in the downlink. On the contrary, the energy consumption is higher and the range is shorter (a range around 5-6 km) due to the high spectrum band used.

The Weightless Special Interest Group developed a set of three open standards for LPWAN: Weightless-W, Weightless-N and Weightless-P.
Weightless-W was developed as a bidirectional (uplink/downlink) solution to operate in TV whitespaces (470-790~MHz).
It is based on narrowband FDMA channels with Time Division Duplex between uplink and downlink; data rate ranges from 1~kbps to 1~Mbps and battery lifetime is around 3-5 years.
Weightless-N was designed to expand the range of Weightless-W and reduce the power consumption (a battery lifetime up to 10 years) at the expense of data rate decrease (from up to 1~Mbps in Weightless-W to 100~kbps in Weightless-N).
Unlike Weightless-W, Weightless-N is based on the Ultra Narrow Band (UNB) technology and operates in the UHF 800-900~MHz band; it provides only uplink communication.
Finally, Weightless-P is proposed as a high-performance two-way communication solution that can operate over 169, 433, 470, 780, 868, 915 and 923~MHz bands.
However, cost of the terminals and power consumption are higher than in Weightless-N, with a battery lifetime of 3-8 years.

Together with \LoRaWAN, SigFox is one of the most adopted LPWAN solutions.
It is a proprietary UNB solution that operates in the 869~MHz (Europe) and 915~MHz (North America) bands.
Its signal is extremely narrowband (100~Hz bandwidth).
It is based on Random Frequency and Time Division Multiple Access (RFTDMA) and achieves a data rate around 100~bps in the uplink, with a maximum packet payload of 12~Bytes, and a number of packets per device that cannot exceed 14~packets/day.
These tough restrictions, together with a business model where SigFox owns the network, have somewhat shifted the interest to~\LoRaWAN, which is considered more flexible and open.

\subsection{Cellular solutions for IoT}

The 3\textsuperscript{rd} Generation Partnership Project (3GPP) standardized a set of low cost and low complexity devices targeting Machine-Type-Communications (MTC) in Release 13. 
In particular, 3GPP addresses the IoT market from a three-fold approach by standardizing the enhanced Machine Type Communications (eMTC), the Narrow Band IoT (NB-IoT) and the EC-GSM-IoT~\cite{nokia16LTEevolution}.

eMTC is an evolution of the work developed in Release 12 that can reach up to 1~Mbps in the uplink and downlink, and operates in LTE bands with a 1.08~MHz bandwidth.
NB-IoT is an alternative that, thanks to the reduced complexity, has a lower cost at the expense of decreasing data rate (up to 250~kbps in both directions).
Finally, EC-GSM-IoT is an evolution of EGPRS towards IoT, with data rate between 70 and 240~kbps.

Although the approaches proposed by 3GPP reduce the energy consumption and the cost of the devices, they have not yet caught up their non-3GPP counterparts. For instance, module cost for \LoRaWAN and SigFox is around \$2-5 and for eMTC is still around \$8-12. Despite the expected broad adoption of cellular IoT solutions supported by 3GPP, \LoRaWAN presents some assets to prevail against these technologies in specific market niches. Current assets are: i) the number of \LoRaWAN network deployments is increasing continuously while, on the other hand, few initial NB-IoT deployments have been already deployed; ii) \LoRaWAN operates in the ISM band whereas cellular IoT operates in licensed bands; this fact favours the deployment of private \LoRaWAN networks without the involvement of mobile operators; iii) \LoRaWAN has backing from industry, e.g. CISCO, IBM or HP, among others. In the future, both technologies will probably coexist when 3GPP solutions will be backed up by large volumes.

\section{Overview of \LoRaWAN}
\label{sec:description}

LoRa is the physical layer used in \LoRaWAN.
It features low power operation (around 10 years of battery lifetime), low data rate (27 kbps with spreading factor 7 and 500 kHz channel or 50 kbps with FSK) and long communication range (2-5 km in urban areas and 15 km in suburban areas).
It was developed by Cycleo (a French company acquired by Semtech).
\LoRaWAN networks are organized in a star-of-stars topology, in which gateway nodes relay messages between end-devices and a central network server.
End-devices send data to gateways over a single wireless hop and gateways are connected to the network server through a non-\LoRaWAN network (e.g.~IP over Cellular or Ethernet). Communication is bi-directional, although uplink communication from end-devices to the network server is strongly favoured, as it will be explained in the following~\cite{sornin16lora}.

\LoRaWAN defines three types of devices (\textit{Class A}, \textit{B} and \textit{C}) with different capabilities~\cite{sornin16lora}.
\textit{Class A} devices use pure ALOHA access for the uplink.
After sending a frame, a \textit{Class A} device listens for a response during two downlink receive windows. Each receive window is defined by the duration, an offset time and a data rate. Although offset time can be configured, the recommended value for each receive window is 1 sec and 2 sec, respectively.
Downlink transmission is only allowed after a successful uplink transmission. The data rate used in the first downlink window is calculated as a function of the uplink data rate and the receive window offset. In the second window the data rate is fixed to the minimum, 0.3 kbps.
Therefore, downlink traffic cannot be transmitted until a successful uplink transmission is decoded by the gateway.
The second receive window is disabled when downlink traffic is received by the end-device in the first window.
\textit{Class A} is the class of \LoRaWAN devices with the lowest power consumption.
\textit{Class B} devices are designed for applications with additional downlink traffic needs.
These devices are synchronized using periodic beacons sent by the gateway to allow the schedule of additional receive windows for downlink traffic without prior successful uplink transmissions.
Obviously, a trade-off between downlink traffic and power consumption arises.
Finally, \textit{Class C} devices are always listening to the channel except when they are transmitting.
Only \textit{class A} must be implemented in all end-devices, and the rest of classes must remain compatible with \textit{Class A}. In turn, \textit{Class C} devices cannot implement \textit{Class B}. The three classes can coexist in the same network and devices can switch from one class to another. However, there is not a specific message defined by \LoRaWAN to inform the gateway about the class of a device and this is up to the application.

The underlying PHY of the three classes is the same.
Communication between end-devices and gateways start with a \textit{Join procedure} that can occur on multiple frequency channels (e.g. in EU863-870 ISM Band there are 3 channels of 125 kHz that must be supported by all end-devices and 3 additional 125 kHz channels) by implementing pseudo-random channel hopping.
Each frame is transmitted with a specific Spreading Factor (SF), defined as $SF= \log_2 {(R_c / R_s)}$, where $R_s$ is the symbol rate and $R_c$ is the chip rate. 
Accordingly, there is a trade-off between SF and communication range. 
The higher the SF (i.e.~the slower the transmission), the longer the communication range.
The codes used in the different SFs are orthogonal.
This means that multiple frames can be exchanged in the network at the same time, as long as each one is sent with one of the six different SFs (from SF=7 to SF=12).
Depending on the SF in use, \LoRaWAN data rate ranges from 0.3~kbps to 27~kbps.

The maximum duty-cycle, defined as the maximum percentage of time during which an end-device can occupy a channel, is a key constraint for networks operating in unlicensed bands. 
Therefore, the selection of the channel must implement pseudo-random channel hopping at each transmission and be compliant with the maximum duty-cycle. For instance, the duty-cycle is 1\% in EU 868 for end-devices.

The LoRa physical layer uses Chirp Spread Spectrum (CSS) modulation, a spread spectrum technique where the signal is modulated by chirp pulses (frequency varying sinusoidal pulses) hence improving resilience and robustness against interference, Doppler effect and multipath.
Packets contain a preamble (typically with 8 symbols), a header (mandatory in explicit mode), the payload (with a maximum size between 51~Bytes and 222~Bytes, depending on the SF) and a Cyclic Redundancy Check (CRC) field (with configurations that provide a coding rate from 4/5 to 4/8).
Typical bandwidth (BW) values are 125, 250 and 500~kHz in the HF ISM 868 and 915 MHz band, while they are 7.8, 10.4, 15.6, 20.8, 31.2, 41.7 and 62.5 kHz in the LF 160 and 480 MHz bands.
The raw data rate varies according to the SF and the bandwidth, and ranges between 22~bps (BW = 7.8~kHz and SF = 12) to 27~kbps (BW = 500~kHz and SF = 7)~\cite{goursaud16dedicated}.
Frequency hopping is exploited at each transmission in order to mitigate external interference~\cite{watteyne09reliability}.

\section{Capacity and Network Size Limitations} 
\label{sec:capacity}

In this section we study the \LoRaWAN network scale with respect to data rate, duty-cycle regulations, etc.

\subsection{Network size limited by duty-cycle}
\label{DutyCycle}

\begin{figure}
    \centering
    \includegraphics[width=1.00\columnwidth]{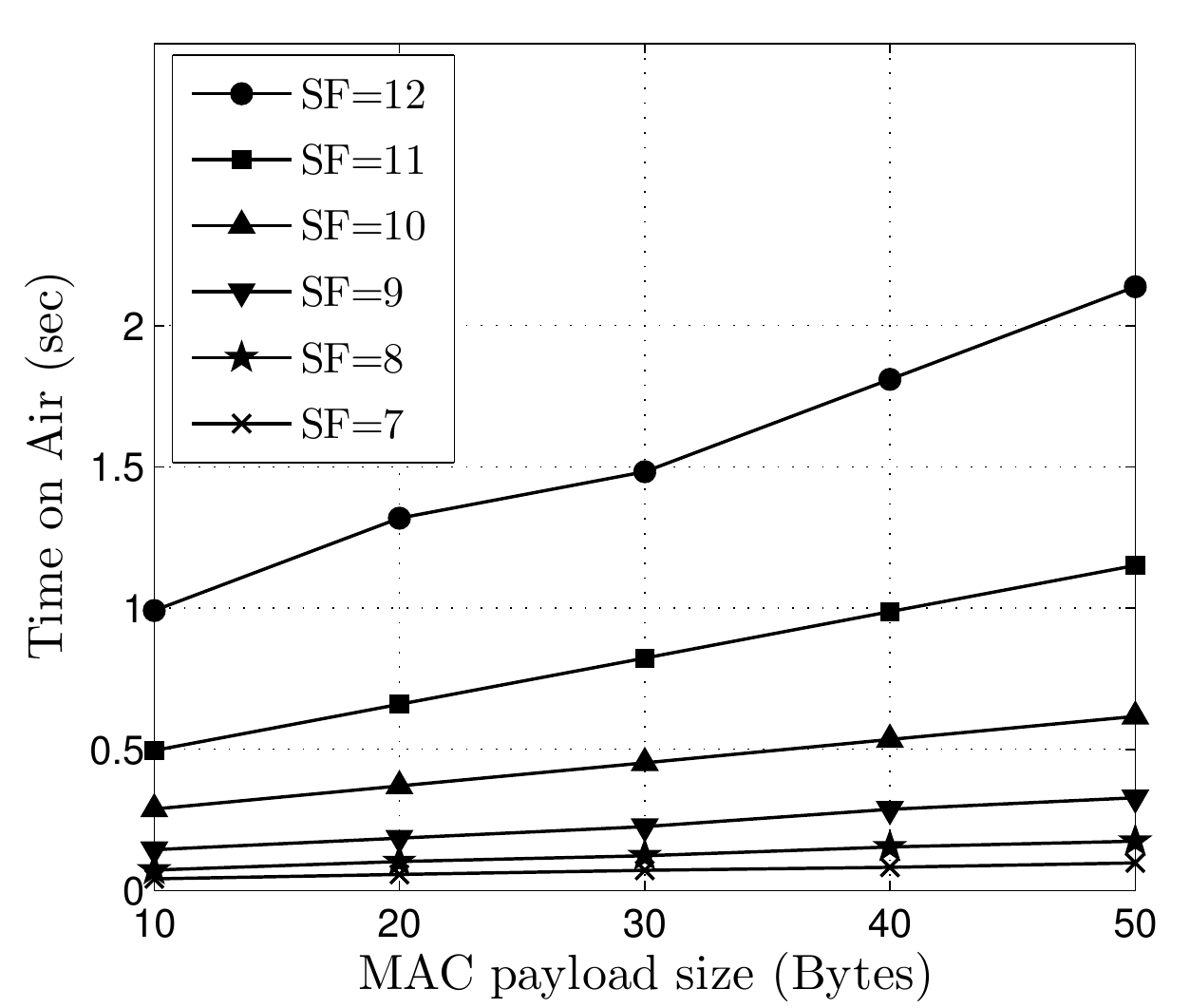}
    \caption{Time on Air of \LoRaWAN with code rate 4/5 and a 125 kHz bandwidth.}
    \label{fig:TimeOnAir}
\end{figure}

Although the performance of \LoRaWAN is determined by PHY/MAC overviewed in Section~\ref{sec:description}, the duty-cycle regulations in the ISM bands~\cite{electronic12erc,federal15fcc} arise as a key limiting factor. If the maximum duty-cycle in a sub-band is denoted by $d$ and the packet transmission time, known as Time On Air, is denoted by $T_a$, each device must be silent in the sub-band for a minimum off-period $T_s= T_a(\frac{1}{d}-1)$. For instance, the maximum duty-cycle of the EU 868 ISM band is 1\% and it results in a maximum transmission time of 36 sec/hour in each sub-band for each end-device. Fig.~\ref{fig:TimeOnAir} shows the Time on Air of a packet transmission with coding rate 4/5 over a 125 kHz bandwidth channel. It is known that large SFs allow longer communication range. However, as observed in Fig.~\ref{fig:TimeOnAir}, large SFs also increase the time on air and, consequently, the off-period duration. This problem is exacerbated by the fact that large SFs are used more often than small SFs. For instance, considering a simple scenario with end-devices distributed uniformly within a round-shaped area centred at the gateway, and a path loss calculated with the Okumura-Hata model for urban cells \cite{Okumura}, the probability that an end-device uses a SF $i$, $p_i$, would be $p_{12}=0.28$, $p_{11}=0.20$, $p_{10}=0.14$, $p_{9}=0.10$, $p_{8}=0.08$ and $p_{7}=0.19$. 

Although Listen Before Talk is not precluded in \LoRaWAN, only ALOHA access is mandatory. Accordingly, the \LoRaWAN capacity can be calculated roughly as the superposition of independent ALOHA-based networks (one independent network for each channel and for each SF, since simultaneous transmissions only cause a collision if they both select the same SF and channel; no capture effect is considered).
However, and in contrast to pure ALOHA, a \LoRaWAN device using SF $i$ cannot exceed a transmitted packet rate given by $nd/T_{a_i}$, where $n$ is the number of channels, $d$ is the duty-cycle and $T_{a_i}$ is the Time On Air with SF $i$.

\begin{figure}
    \centering
    \includegraphics[width=0.9\columnwidth]{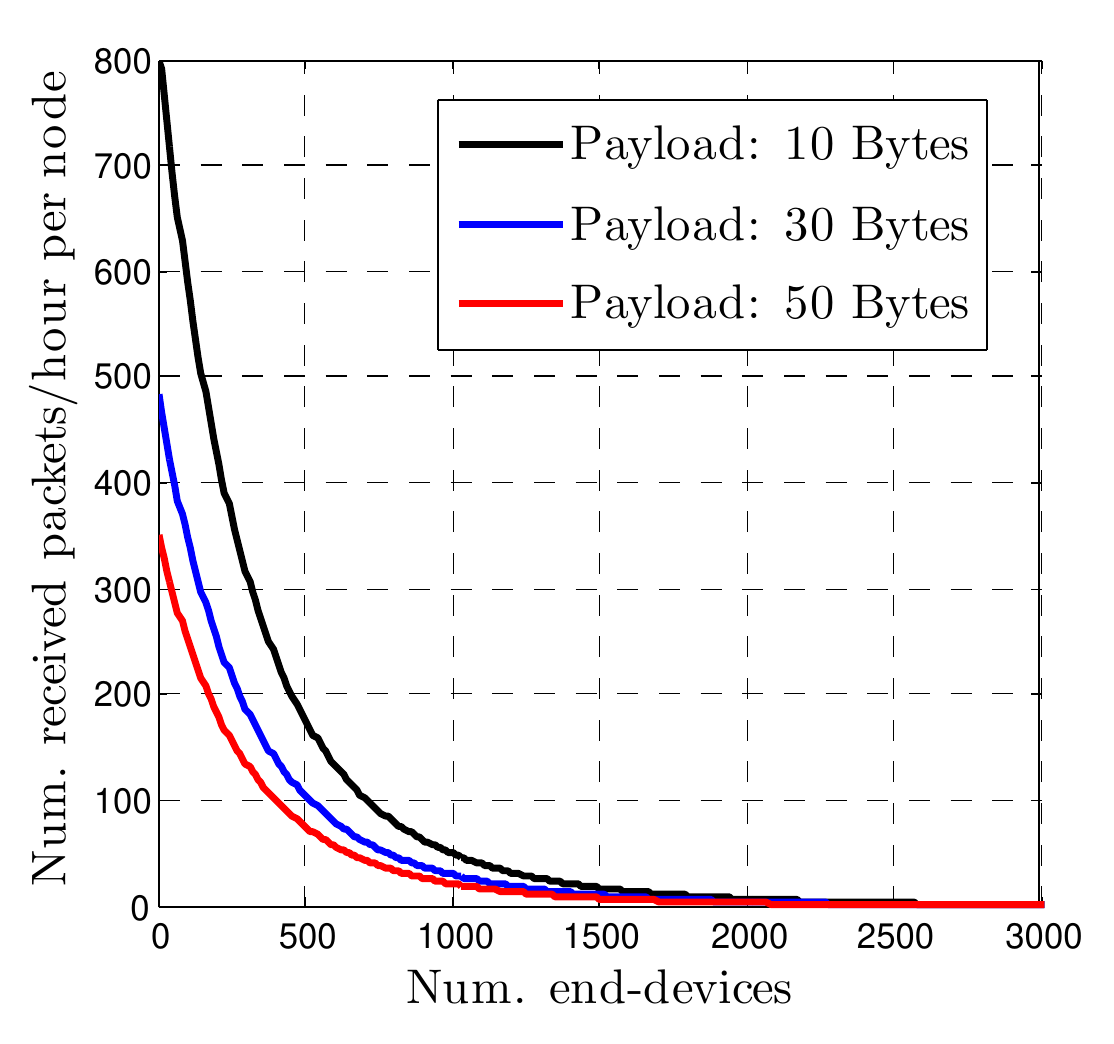}
    \caption{Number of packets received per hour when end-devices attempt transmission at $nd/T_{a_i}$ packets/sec with coding rate 4/5 and $n=3$ channels with 125 kHz bandwidth.}
    \label{fig:PacksDC}
\end{figure}

In the simple scenario described above, if all end-devices transmit packets at the maximum packet rate $nd/T_{a_i}$, the number of packets successfully received by the gateway decreases as shown in Fig.~\ref{fig:PacksDC}, where a network with $n=3$ channels has been analyzed.
The number of received packets drops due to the effect of collisions.

\begin{figure}
    \centering
    \includegraphics[width=0.9\columnwidth]{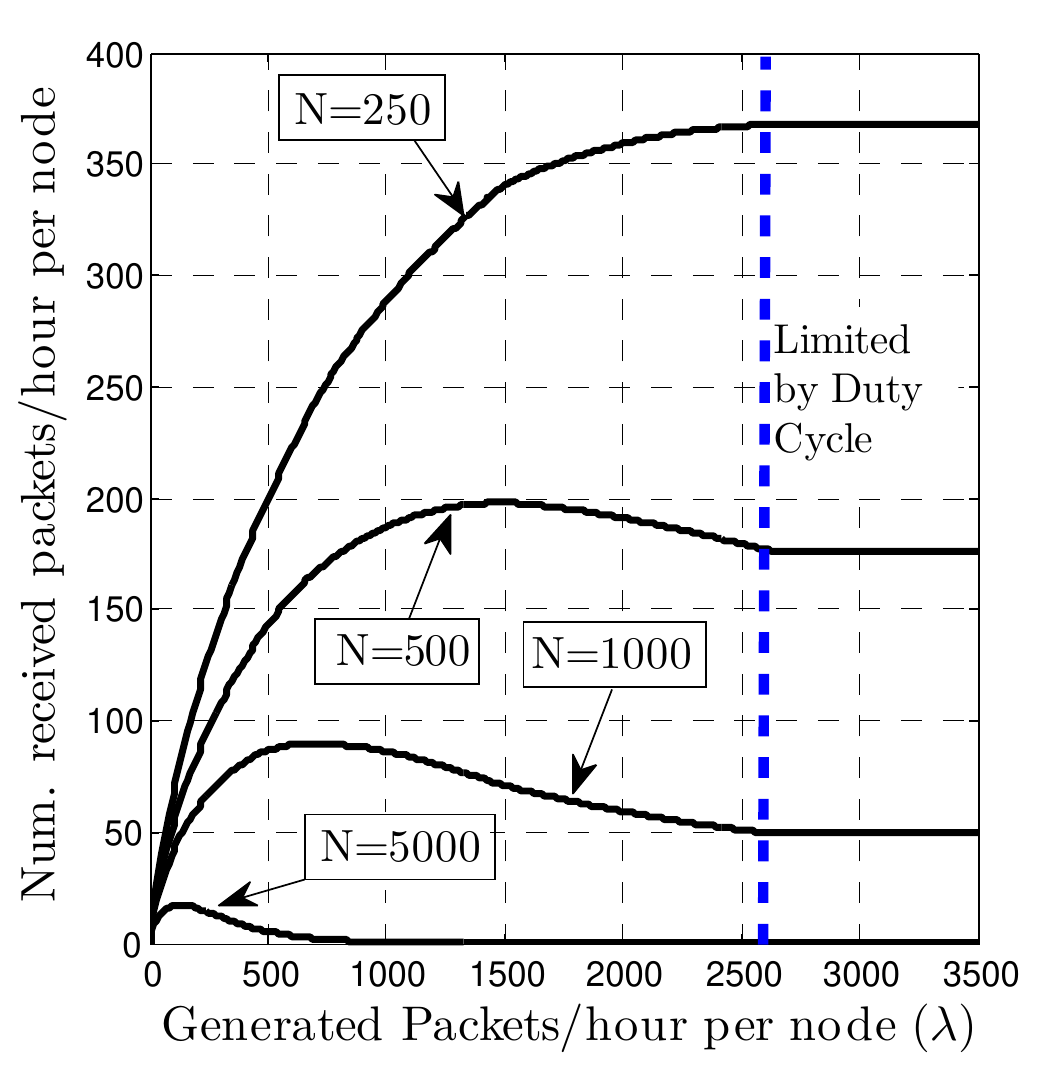}
    \caption{Number of 10 Bytes payload packets received per hour and node for $\{  250, 500, 1000, 5000\}$ end-devices and $n=3$ channels as a function of the packet generation.}
    \label{fig:comparison}
\end{figure}

In Fig.~\ref{fig:comparison} the number of packets received successfully  per hour and end-device is shown for deployments with $\{  250, 500, 1000, 5000\}$ end-devices and $n=3$ channels.
For low transmission rate values (in packets/hour), throughput is limited by collisions; for high values, the maximum duty-cycle prevents end-devices from increasing the packet transmission rate and stabilizes the throughput.
For deployments with a ``small'' number of end-devices, the duty-cycle constraint limits the maximum throughput.

Table~\ref{t:throughput} summarizes the maximum throughput per end-device and the probability of successful reception for a set of different deployments.
The maximum throughput falls as the number of end-devices grows. 

\begin{table*}
    \centering
    \caption{Maximum throughput and probability of successful transmission for different deployments (with $n$=3 channels and 1\% duty-cycle)}
    \label{t:throughput}
    \begin{tabular}{l|ccc|ccc|ccc|ccc|}
                                                                                \cline{2-13}
                                                                              & \multicolumn{3}{c|}{250 end-devices}                           & \multicolumn{3}{c|}{500 end-devices}                          & \multicolumn{3}{c|}{1000 end-devices}                          & \multicolumn{3}{c|}{5000 end-devices}                         \\
        \hline
        \multicolumn{1}{|l|}{Payload (Bytes)}                                 & \multicolumn{1}{c|}{10} & \multicolumn{1}{c|}{30} & 50 & \multicolumn{1}{c|}{10} & \multicolumn{1}{c|}{30} & 50 & \multicolumn{1}{c|}{10} & \multicolumn{1}{c|}{30} & 50 & \multicolumn{1}{c|}{10} & \multicolumn{1}{c|}{30} & 50 \\
        \hline
        \multicolumn{1}{|l|}{Max. throughput per node (Packets/hour)}         &   367 &    217 &    157                                  &    198 &    117 &   84                                  &    89 &    53 &    38                                  &   18 &   10 &   7.3                                  \\
        \multicolumn{1}{|l|}{Max. throughput per node (Bytes/hour)}           &  3670 &  6510 &  7850                                  &   1980 &  3510 &  4200                                  &   890 &   1590 &   1900                                  &    180 &   300 &   365                                  \\
        \multicolumn{1}{|l|}{$\lambda$ of the max. throughput (Packets/hour)} &   2620 &   1500 &   1090                                  &   1500 &   870 &   620                                  &   670 &    390 &    280                                  &    130 &    70 &  50                                  \\ 
        \multicolumn{1}{|l|}{Prob. of successful transmission (\%)}           & 14.01 & 14.47 & 10.73                                  & 13.20 & 13.45 & 13.55                                  & 13.28 & 13.59 & 13.57                                  & 13.85 & 14.29 & 14.60                                  \\
        \hline
\end{tabular}
\end{table*}

\subsection{Reliability and Densification drain Network Capacity}

In \LoRaWAN, reliability is achieved through the acknowledgment of frames in the downlink.
For class A end-devices, the acknowledgment can be transmitted in one of the two available receive windows; for class B end-devices it is transmitted in one of the two receive windows or in an additional time-synchronized window; for class C end-devices it can be transmitted at any time .

In \LoRaWAN the capacity of the network is reduced not only due to transmissions in the downlink, but also due to the off-period time following those transmissions (gateways must be compliant with duty-cycle regulation).
Therefore, the design of the network and the applications that run on it must minimize the number of acknowledged frames to avoid the capacity drain.
This side-effect calls into question the feasibility of deploying ultra-reliable services over large-scale \LoRaWAN networks.

At this point of development of the technology, \LoRaWAN faces deployment trends that  can result in future inefficiencies.
Specifically, \LoRaWAN networks are being deployed following the cellular network model, that is, network operators provide connectivity as a service.
This model is making gateways to become base stations covering large areas. The increase in the number of end-devices running applications from different vendors over the same shared infrastructure poses new challenges to coordinate the applications. In particular, each application has specific constraints in terms of reliability, maximum latency, transmission pattern, etc. The coordination of the diverse requirements over a single shared infrastructure using an ALOHA-based access is one of the main future challenges for the technology. Therefore, a fair spectrum sharing is required beyond the existing duty-cycle regulations.     
Finally, the unplanned and uncoordinated deployment of \LoRaWAN gateways in urban regions, along with the deployment of alternative LPWAN solutions (e.g. SigFox), could cause a decrease of the capacity due to collisions and due to the use of larger SFs (to cope with higher interference levels).

\section{Use Cases}
\label{sec:usecases}

Several application use cases are considered in order to analyze the suitability of \LoRaWAN and complement the understanding of the advantages and limitations of the technology when applied to different types of data transmission patterns, latency requirements, scale and geographic dispersion among others.

\subsection{Real Time Monitoring}

Agriculture, leak detection or environment control are applications with a reduced number of periodic/aperiodic messages and relaxed delay constraints. In contrast, the communication range must be long enough to cope with dispersed location of end-devices. \LoRaWAN has been designed to handle the traffic generated by this type of applications and meets their requirements as long as the deployment of the gateways is enough to cover all end-devices.

On the other hand, industrial automation, critical infrastructure monitoring and actuation require some sort of real time operation.
Real time is understood in general by low latency, and bounded jitter and depends on the specific application.
\LoRaWAN technology cannot claim to be a candidate solution for industrial automation, considering for example that industrial control loops may require response times around $1$ ms to $100$ ms and that, even for small packets of 10 Bytes, the time on air with SF=7 is around 40 ms.
As presented in the previous section, due to the MAC nature of \LoRaWAN, deterministic operation cannot be guaranteed despite of application specific periodicity as ALOHA access is subject to contention which impacts network jitter.
Despite that, small \LoRaWAN networks can deliver proper service to applications that require, for instance, sampling data every second.
To do that, two main design considerations should be taken into account:

\begin{itemize}
    \item The spreading factor should be as small as possible to limit both the time on air and the subsequent off-period.
        In other words, the gateway must be close enough to the end-devices.
    \item The number of channels must be carefully designed and must be enough to i) minimize the probability of collisions (tightly coupled with the number of end-devices) and ii) offer quick alternative channels for nodes to retransmit collided packets thereby diminishing the impact of the duty-cycle.
\end{itemize}

Despite the two aforementioned aspects, latency will not be deterministic.

\subsection{Metering}

The \LoRa Alliance is working on standard encapsulation profiles for popular M2M and metering protocols.
Keeping an existing application layer allows to keep intact most of the firmware and ecosystem, facilitating migration to LPWAN.
These protocols include Wireless M-Bus for water or gas metering, KNX for building automation, and ModBus for industrial automation.
It is important to understand that those scenarios range from time sensitive operation to best effort monitoring.
Therefore, it is key to identify in such a diverse ecosystem what the requirements of each application are and if \LoRaWAN is the appropriate technology to address them.

\subsection{Smart City Applications}

\LoRaWAN has shown key success stories with smart lighting, smart parking and smart waste collection thanks to their scale and the nature of the data generated by those applications.
These encompass periodic messaging with certain delay tolerance.
For example, smart parking applications report the status of the parking spots upon a change is detected~\cite{martinez15lean}.
Parking events are slow and therefore network signaling is limited to few tens of messages per day.
Analogously smart waste collection systems and smart lighting actuate or report information in response to a measure with large variation periods.
Although latency and jitter are not major issues in these applications, in some of them the triggering factor is simultaneous for a huge number of end-devices.
For instance, sunset and down trigger the lighting elements around the whole city, thereby causing an avalanche of messages.
\LoRaWAN is an appropriate technology for this use case since it handles the wide coverage area and the significant number of users at the expense of increasing number of collisions, latency and jitter.

\subsection{Smart Transportation and Logistics}

Transportation and logistics are seen as two major pillars of the expected IoT growth over the next few years thanks to their impact on the global economy. Most applications are targeting efficiency in areas such as public transportation or transport of goods. However, some applications are tolerant to delay, jitter or unreliability and some others are not.

Different standards have been developed in the 5.9 GHz band for Intelligent Transportation Systems (ITS) based on the IEEE 802.11p standard. The constraints on delay are diverse for different applications, but \LoRaWAN, being a LPWAN solution, is not suitable for these applications. On the contrary, solutions such as fleet control and management can be supported by \LoRaWAN. Roaming is one of the developments under definition within \LoRa Alliance to enhance mobility. Specifically, future roaming solution is expected to support back-end to back-end secure connections, clearing and billing between operators, location of end-devices (pointed out as an open research challenge in Section \ref{sec:research}) and transparent device provisioning across networks.


\subsection{Video Surveillance}

The most common digital video formats for IP-based video systems are MJPEG, MPEG-4 and H.264. The bit rate recommended for IP surveillance cameras ranges from 130 kbps with low quality MJPEG coding to 4 Mbps for 1920x1080 resolution and 30 fps MPEG-4/H.264 coding. Given that \LoRaWAN data rate ranges from 0.3~kbps to 50~kbps per channel, \LoRaWAN will not support these applications.

\section{Open Research Challenges}
\label{sec:research}

The effect of the duty-cycle stated in Section~\ref{sec:capacity} jeopardizes the actual capacity of large-scale deployments.
This has been initially addressed by TheThingsNetwork~\cite{giezeman16things}, an interesting global, open, crowd-sourced initiative to create an Internet of Things data network over \LoRaWAN technology.
The proposed solution defines an access policy, known as the TTN Fair Access Policy, that limits the Time on Air of each end-device to a maximum of 30 sec per day.
This policy is simple to implement and guarantees pre-defined end-device requirements for a large-scale network (more than 1000 end-devices per gateway).
However, it fails to provide the network with enough flexibility to adapt to environment and network conditions (i.e.~link budget of each end-device, number of end-devices, number of gateways, etc), as well as to applications with tight latency or capacity requirements.

At this stage, the optimization of the capacity of the \LoRaWAN network, as well as the possibility to perform traffic slicing for guaranteeing specific requirements in a service basis, remain as open research issues.
From the authors' point of view, the research community will have to address the following open research challenges during the next years:           

\begin{itemize}
    \item \textbf{Explore new channel hopping methods:} 
    A pseudo-random channel hopping method is natively used in \LoRaWAN to distribute transmissions over the pool of available channels, thereby reducing the collision probability. However, this method cannot meet traffic requirements when there are latency, jitter or reliability constraints (i.e. downlink ACKs for all packets), and it is not able to get adapted according to the noise level of each channel.
    
    The design of pre-defined and adaptive hopping sequences arises as an open research issue. From the authors' point of view, the proposed channel hopping sequences should be able to reserve a set of channels for retransmissions of critical packets, both in the uplink and in the downlink (ACK). 
    
    The design of feasible feedback mechanisms between gateways and end-devices must be a key part of the approach in a system where uplink traffic is strongly favoured. 
    
    \item \textbf{Time Division Multiple Access (TDMA) over \LoRaWAN:} 
    The random nature of ALOHA-based access is not optimal to serve deterministic traffic, which is gaining importance in the IoT ecosystem. Building a complete or hybrid TDMA access on top of \LoRaWAN opens up new use cases for this technology and provides additional flexibility.
    
    The TDMA scheduler should be able to allocate resources for ALOHA-based access and schedule deterministic traffic along time and over the set of available channels. 
The proposed schedulers should manage the trade-off between resources devoted for deterministic and non-deterministic traffic, meet the regional duty-cycle constraints and guarantee fairness with co-existing \LoRaWAN  networks.

    \item \textbf{Geolocation of end-devices:} 
    The location of end-devices is a mandatory requirement for specific use cases, particularly in industry 4.0. However, GPS-based solutions are not feasible due to cost, and CPU and energy consumption. Currently, interesting works have been initiated to develop TDOA-based (Time Difference Of Arrival) triangulation techniques for \LoRaWAN. It has been shown that this approach benefits from large SFs and dense gateway deployments.  

    \item \textbf{Cognitive Radio:}
    As pointed out in Section \ref{DutyCycle}, regulation in ISM bands concerning maximum duty-cycle has a significant impact on the capacity of the network. One of the most promising future directions could be the inclusion of cognitive radio into the \LoRaWAN standard. In contrast to Weightless-W, \LoRaWAN has not been designed to operate in TV whitespaces. In the future, the inclusion of cognitive radio into the \LoRaWAN standard would be subject to a significant reduction of the energy consumption associated with cognitive radio techniques.     

    \item \textbf{Power reduction for multi-hop solutions:}
    \LoRaWAN is organized with a single-hop star topology for simplicity. As discussed in Section \ref{sec:capacity}, the impact of high SFs on the capacity of the network is two-fold, since it increases both the Time on Air and the off-period. A two-hop strategy for \LoRaWAN networks should be investigated to figure out its potential.
    
    Proposals in this direction should consider the reduction of transmitted power and the decrease of the SFs. On the other hand, also negative effects such as complexity, synchronization, and increasing power consumption of relays should be analyzed to thoroughly characterize the trade-off.  

\item \textbf{Densification of \LoRaWAN networks:}
The proliferation of LPWAN technologies, and particularly  \LoRaWAN, poses co-existence challenges as the deployment of gateways populate urban areas. Given the random-based access in unlicensed bands of \LoRaWAN and its inherent unplanned deployment, the performance achieved in isolated networks is put into question in scenarios with co-existing gateways and limited number of available channels.

It is essential to devise coordination mechanisms between gateways from the same or different operators to limit interference and collisions. The co-existence mechanisms encompass coordination and reconfiguration protocols for gateways and end-devices.
\end{itemize}

\section{Conclusions}
\label{sec:conclusion}

This article is aimed to clarify the scope of \LoRaWAN by exploring the limits of the technology, matching them to application use cases and stating the open research challenges.
In the low power M2M fragmented connectivity space there is not a single solution for all the possible connectivity needs and \LoRaWAN is not an exception.
A \LoRaWAN gateway, covering a range of tens of kilometers and able to serve up to thousands of end-devices, must be carefully dimensioned to meet the requirements of each use case.
Thus, the combination of the number of end-devices, the selected SFs and the number of channels will determine if the \LoRaWAN ALOHA based access and the maximum duty-cycle regulation fit each use case.
For instance, we have seen that deterministic monitoring and real time operation cannot be guaranteed with current \LoRaWAN state of the art.

\section*{Acknowledgment}

This work is partially supported by the Spanish Ministry of Economy and the FEDER regional development fund under SINERGIA project (TEC2015-71303-R), and by the European Commission through projects H2020~F-Interop and H2020~ARMOUR.

\newpage
\bibliographystyle{IEEEtran}


\begin{IEEEbiographynophoto}{Ferran Adelantado}
(ferranadelantado@uoc.edu) received the Engineering degree in Telecommunications (2007) and the PhD degree in Telecommunications (2007) from UPC, and the BSc in Business Sciences (2012) from UOC. Currently, he is associate professor at Universitat Oberta de Catalunya (UOC) and researcher at the Wireless Networks Research Group (WINE). His research interests are wireless networks, particularly 5G, LPWAN and IoT technologies.
\end{IEEEbiographynophoto}

\begin{IEEEbiographynophoto}{Xavier Vilajosana}
is Principal Investigator of the Wireless Networks Research Lab at the Open University of Catalonia. Xavier is also co-founder of Worldsensing and OpenMote Technologies. Xavier is an active member of the IETF 6TiSCH WG where he authored different standard proposals. Xavier also holds 30 patents. Xavier has been visiting professor at the Prof. Pister UC Berkeley lab. He co-leads Berkeley's OpenWSN project. Xavier has been Senior Researcher at the HP R\&D labs (2014-2016) and visiting researcher at the France Telecom R\&D Labs Paris (2008). He holds a PhD(2009), MSc and MEng (2004) from UPC, Barcelona, Spain.
\end{IEEEbiographynophoto}

\begin{IEEEbiographynophoto}{Pere Tuset-Peiro}
[M'12] (peretuset@uoc.edu) is Assistant Professor at the Department of Computer Science, Multimedia and Telecommunications and researcher at the Internet Interdisciplinary Institute (IN3), both of Universitat Oberta de Catalunya (UOC). He received the BSc and MSc in Telecommunications Engineering from Universitat Polit\`ecnica de Catalunya (UPC) in 2007 and 2011 respectively, and the PhD in Network and Information Technologies from UOC in 2015. Currently, he holds more than 20 high-impact publications and 7 international patents.
\end{IEEEbiographynophoto}

\begin{IEEEbiographynophoto}{Borja Martinez}
received the B.Sc. in Physics and Electronics Engineering, the M.Sc. in Microelectronics and the Ph.D. in Computer Science from the Universidad Aut\'onoma de Barcelona (UAB), Spain. From 2005 to 2015 he was assistant professor at the Department of Microelectronics and Electronic Systems of the UAB. He is currently a research fellow at the Internet Interdisciplinary Institute (IN3-UOC). His research interests include low-power techniques for smart wireless devices, energy efficiency and algorithms.
\end{IEEEbiographynophoto}

\begin{IEEEbiographynophoto}{Joan Meli\`a-Segu\'i}
(melia@uoc.edu), Ph.D. (2011), is a lecturer at the Estudis de Inform\`atica, Multim\`edia i Telecomunicaci\'o and a researcher at the Internet Interdisciplinary Institute, both at Universitat Oberta de Catalunya. Before, he was a postdoctoral researcher at Universitat Pompeu Fabra and the Palo Alto Research Centre (Xerox PARC). He has published more than 30 papers and one patent in the areas of the Internet of Things, intelligent systems, security, and privacy.
\end{IEEEbiographynophoto}

\begin{IEEEbiographynophoto}{Thomas Watteyne}
(www.thomaswatteyne.com) is a researcher in the EVA team at Inria in Paris, and a Senior Networking Design Engineer at Linear Technology/Dust Networks in Silicon Valley. He co-chairs the IETF 6TiSCH working group. Thomas did his postdoctoral research with Prof. Pister at UC Berkeley. He co-leads Berkeley's OpenWSN project. In 2005-2008, he was a research engineer at France Telecom, Orange Labs. He holds a PhD (2008), MSc and MEng (2005) from INSA Lyon, France.
\end{IEEEbiographynophoto}

\end{document}